\documentclass[conference,letterpaper]{IEEEtran}

\usepackage{perlaza}
\usepackage{tikz}

\newcommand{\ts}{\textsuperscript}
\makeatletter
\DeclareFontFamily{U}{tipa}{}
\DeclareFontShape{U}{tipa}{m}{n}{<->tipa10}{}
\newcommand{\arc@char}{{\usefont{U}{tipa}{m}{n}\symbol{62}}}%

\DeclareMathOperator*{\argm}{arg\,min}

\newcommand{\arc}[1]{\mathpalette\arc@arc{#1}}

\newcommand{\arc@arc}[2]{%
  \sbox0{$\m@th#1#2$}%
  \vbox{
    \hbox{\resizebox{\wd0}{\height}{\arc@char}}
    \nointerlineskip
    \box0
  }%
}
\makeatother
\makeatletter

\IEEEoverridecommandlockouts
\begin{document}
\title{Achievable Information-Energy Region in the Finite Block-Length Regime with Finite Constellations} 



\author{Sadaf ul Zuhra, Samir M. Perlaza, H. Vincent Poor, and Eitan Altman
		\thanks{Sadaf ul Zuhra, Samir M. Perlaza, and Eitan Altman are with INRIA, Centre de Recherche de Sophia Antipolis - M\'{e}diterran\'{e}e, 2004  Route des Lucioles, 06902 Sophia Antipolis, France. $\lbrace$sadaf-ul.zuhra, samir.perlaza, eitan.altman$\rbrace$@inria.fr}
		\thanks{H. Vincent Poor (poor@princeton.edu) and Samir M. Perlaza are with the Department of Electrical and Computer Engineering, Princeton University, Princeton, 08544 NJ, USA.}
		\thanks{Eitan Altman is also with the Laboratoire d'Informatique d'Avignon (LIA), Universit\'{e} d'Avignon, 84911 Avignon, France; and with the Laboratory of Information, Network and Communication Sciences (LINCS), 75013 Paris, France.}
		\thanks{Samir M. Perlaza is also with the Laboratoire de Math\'{e}matiques GAATI, Universit\'{e} de la Polyn\'{e}sie Fran\c{c}aise,  BP 6570, 98702 Faaa, French Polynesia.}
  \thanks{This research was supported in part by the European Commission through the H2020-MSCA-RISE-2019 program under grant 872172; in part by the Agence Nationale de la Recherche (ANR) through the project MAESTRO-5G (ANR-18-CE25-0012); in part by the U.S. National Science Foundation under Grant CCF-1908308; and in part by the French Government through the ``Plan de Relance" and ``Programme d’investissements d’avenir".}
	}
\maketitle

\begin{abstract}
This paper characterizes an achievable information-energy region of simultaneous information and energy transmission over an additive white Gaussian noise channel. This analysis is performed in the finite block-length regime with finite constellations. More specifically, a method for constructing a family of codes is proposed and the set of achievable tuples of information rate, energy rate, decoding error probability (DEP) and energy outage probability (EOP) is characterized. Using existing converse results, it is shown that the construction is information rate, energy rate, and EOP optimal. The achieved DEP is, however, sub-optimal.
\end{abstract}

\setlength{\textfloatsep}{1pt}



\section{Introduction} \label{SecIntro}
Simultaneous information and energy transmission (SIET), also known as simultaneous wireless information and power transfer (SWIPT) is one of the key technologies being researched~\cite{varshney2008transporting, amor2016fundamental, perlaza2018simultaneous,NizarItw2021,6373669} for use in 6G systems~\cite{6GSpeculation}. A key research direction is the study of the fundamental trade-off between the amount of information and energy that can be simultaneously transmitted by a signal. This trade-off has been studied in~\cite{varshney2008transporting,GroverSahai} in the case of point-to-point noisy channels in the asymptotic regime. Nonetheless, the information-energy trade-off is not the only trade-off involved in SIET. In the finite block-length regime, several other trade-offs appear which are taken into consideration in this paper.

Optimal signal and system design exclusively for wireless energy transmission has been studied in~\cite{9447959,7547357,9411899,9184149,7867826, 9447237, 9153166}. Various aspects of SIET system design such as resource allocation, receiver architectures, energy harvester circuits, and decoding strategies have been considered in~\cite{9241856,9149424,9377479,6898012, 9295579,7998252}. An algorithm for designing circular quadrature amplitude modulation for SIET that maximizes the peak-to-average power ratio has been proposed in~\cite{9593249}.

Much of the existing work in this field including~\cite{varshney2008transporting,amor2016fundamental,NizarItw2021,GroverSahai,amor2016feedback,KhalfetGIC} assume that the duration of transmission is infinitely long. The assumption of infinitely long transmissions guarantees that the decoding error probability (DEP) and the energy outage probability (EOP) can be made arbitrarily close to zero, and thus, the focus is only on the information and energy transmission rates. In the finite block-length regime, which is the subject of this paper, the DEP and EOP are bounded away from zero. 

Earlier research on the fundamental limits of SIET in the finite block-length regime is presented in~\cite{perlaza2018simultaneous,khalfet2019ultra,zuhraITW}. In~\cite{perlaza2018simultaneous} and~\cite{khalfet2019ultra}, the information-energy capacity region with binary antipodal channel inputs is presented. A converse information-energy region of SIET with arbitrary number of channel inputs is presented in~\cite{zuhraITW}.
Comprehensive overviews of the literature on SIET can be found in~\cite{survey,surveyA,8476597,ClerckxFoundations}.

This work proposes a new method of constructing a family of codes for SIET over an AWGN channel in the finite block-length regime with finite constellation sizes. An achievable information-energy region is characterized for the constructed family of codes. Using the converse results obtained in part from~\cite{zuhraITW}, it is shown that the constructed codes are information rate, energy rate and EOP optimal.


The rest of this paper is organized as follows. Section~\ref{SecConstruction} presents a new method for constructing families of codes that satisfy the information and energy rate, DEP, and EOP requirements. The converse and achievable information-energy regions are characterized in Section~\ref{SecConverse} and Section~\ref{SecMain}, respectively. The gap between the converse and achievable regions is studied using numerical examples in Section~\ref{SecExample}. 

\section{System Model} \label{SecSystemModel}
Consider a communication system formed by a transmitter, an information receiver (IR), and an energy harvester (EH). The objective of the transmitter is to simultaneously send information to the IR at a rate of $R$ bits per second; and energy to the EH at a rate of $B$ Joules per second over an AWGN channel. 
That is, given a channel input
${\boldsymbol x} = (x_1,x_2, \ldots, x_n)^{\sf{T}} \in \mathds{C}^{n}$, with $n \in \mathds{N}$, the outputs of the channel are the random vectors
\begin{subequations}\label{EqChannelModel}
\begin{IEEEeqnarray}{rcl}
    \label{eq:channel1}
    {\boldsymbol Y} & = & {\boldsymbol x} + {\boldsymbol N}_1, \mbox{ and }  \\
    \label{eq:channel2} 
    {\boldsymbol Z} & = & {\boldsymbol x} + {\boldsymbol N}_2,
\end{IEEEeqnarray}
\end{subequations}
where $n$ is the duration of the transmission in channel uses; and the vectors ${\boldsymbol Y} = (Y_1,Y_2, \ldots, Y_n)^{\sf{T}} \in \mathds{C}^{n}$ and ${\boldsymbol Z} = (Z_1,Z_2, \ldots, Z_n)^{\sf{T}} \in \mathds{C}^{n}$ are the inputs of the IR and the EH, respectively. The components of the random vectors ${\boldsymbol N}_1 = (N_{1,1}, N_{1,2}, \ldots, N_{1,n})^{\sf{T}}\in \mathds{C}^{n}$ and ${\boldsymbol N}_2 = (N_{2,1}, N_{2,2}$, $\ldots$, $N_{2,n})^{\sf{T}}$ $\in$ $\mathds{C}^{n}$ are independent and identically distributed. More precisely, for all $(i,j) \in \{1,2\} \times \{1,2, \ldots, n\}$,  $N_{i,j}$ is a complex circularly symmetric Gaussian random variable whose real and imaginary parts have zero means and variances $\frac{1}{2}\sigma^2$. 
%
That is, for all $\boldsymbol{y} = (y_1, y_2, \ldots, y_n)^{\sf{T}} \in \mathds{C}^{n}$, for all $\boldsymbol{z} = (z_1, z_2, \ldots, z_n)^{\sf{T}} \in \mathds{C}^{n}$, and for all $\boldsymbol{x} = (x_1, x_2, \ldots, x_n)^{\sf{T}} \in \mathds{C}^{n}$, it holds that the joint probability density function of the channel outputs $(\boldsymbol{Y}, \boldsymbol{Z})$ satisfies $f_{\boldsymbol{Y Z}|\boldsymbol{X}}(\boldsymbol{y}, \boldsymbol{z}|\boldsymbol{x}) = f_{\boldsymbol{Y}|\boldsymbol{X}}(\boldsymbol{y}|\boldsymbol{x}) f_{\boldsymbol{Z}|\boldsymbol{X}}(\boldsymbol{z}|\boldsymbol{x})$, where  
\begin{IEEEeqnarray}{rcl} \label{EqYXdistribution}
    f_{\boldsymbol{Y}|\boldsymbol{X}}(\boldsymbol{y}|\boldsymbol{x}) & = &  \prod_{t=1}^n f_{Y|X}(y_t|x_t) \mbox{ and }\\
    \label{eq:z_distribution}
    f_{\boldsymbol{Z}|\boldsymbol{X}}(\boldsymbol{z}|\boldsymbol{x}) & = & \prod_{t=1}^n f_{Z|X}(z_t|x_t),
\end{IEEEeqnarray}
and for all $t \in \lbrace 1,2, \ldots, n \rbrace$,
\begin{subequations}\label{EqDensities}
\begin{IEEEeqnarray}{rcl}
\label{EqYXdistribution2}
&&f_{Y|X}(y_t|x_t) = \frac{1}{\pi \sigma^2}\exp \left( - \frac{\left| y_t - x_t \right|^2}{\sigma^2} \right), \\
%
\label{eq:z_distribution2}
 &&f_{Z|X}(z_t|x_t) = \frac{1}{\pi \sigma^2}\exp \left( - \frac{\left| z_t - x_t \right|^2}{\sigma^2} \right).
\end{IEEEeqnarray}
\end{subequations}
Within this framework, the two tasks of information and energy transmission must be accomplished. 

\subsection{Information Transmission} \label{subsec:information_transmission}
Assume that the information transmission takes place using a modulation scheme that uses $L$ symbols. That is, there is a set
\begin{equation}\label{EqCIsymbols}
\mathcal{X} \triangleq \{x^{(1)}, x^{(2)}, \ldots, x^{(L)}\} \subset \mathds{C}
\end{equation} that contains all possible channel input symbols, and 
\begin{equation}\label{EqL}
L \triangleq \left| \mathcal{X} \right|.
\end{equation}
Let $M$ be the number of message indices to be transmitted within $n$ channel uses. That is,
\begin{equation} \label{eq:M_L}
  M \leqslant L^n.  
\end{equation}
To reliably transmit a message index, the transmitter uses an $(n,M)$-code defined as follows.
\begin{definition}[$(n,M)$-code] 
\label{DefNmCode}
An $(n,M)$-code for the random transformation in~\eqref{EqChannelModel} is a system
\begin{equation}
    \left \lbrace ({\boldsymbol u}(1),\mathcal{D}_1), ({\boldsymbol u}(2),\mathcal{D}_2), \ldots, ({\boldsymbol u}(M),\mathcal{D}_M)\right \rbrace,
\end{equation}
where, for all $(i,j) \in \{1,2, \ldots, M\}^2, i\neq j$,
\begin{subequations}\label{EqCodeProperties}
\begin{align}
\label{eq:u_i}  &{\boldsymbol u}(i) = (u_1(i), u_2(i), \ldots, u_n(i)) \in \mathcal{X}^n, \\
        &\mathcal{D}_i \cap \mathcal{D}_j = \phi,\\
        &\bigcup_{i = 1}^M \mathcal{D}_i \subseteq \mathds{C}^n, \mbox{ and }\\
  \label{EqPCriteria}      &|u_t(i)| \leqslant P,
    \end{align}
\end{subequations}
\end{definition}
where $P$ is a peak-power constraint.
Assume that the transmitter uses an $(n,M)$-code
\begin{equation} \label{Eqnm_code}
    \mathscr{C} \triangleq \{({\boldsymbol u}(1),\mathcal{D}_1), ({\boldsymbol u}(2),\mathcal{D}_2), \ldots, ({\boldsymbol u}(M),\mathcal{D}_M)\},
\end{equation}
that satisfies~\eqref{EqCodeProperties}. Without any loss of generality, assume that for all $i \inCountK{M}$, the decoding set $\mathcal{D}_i$ is written in the form
\begin{equation}\label{EqDit}
    \mathcal{D}_i = \mathcal{D}_{i,1} \times \mathcal{D}_{i,2} \times \ldots \times \mathcal{D}_{i,n},
\end{equation}
where for all $t \inCountK{n}$, the set $\mathcal{D}_{i,t}$ is a subset of~$\complex$.
The information rate of any $(n,M)$-code $\mathscr{C}$ is given by
\begin{equation} \label{EqR}
    R = \frac{\log M}{n}
\end{equation}
in bits per channel use.
To transmit message index $i$, the transmitter uses the codeword ${\boldsymbol u}(i)= (u_1(i), u_2(i), \ldots, u_n(i))$. That is, at channel use $t$, the transmitter inputs the symbol $u_{t}(i)$ into the channel. At the end of $n$ channel uses, the IR observes a realization of the random vector ${\boldsymbol Y} = (Y_1, Y_2, \ldots, Y_n)^{\sf{T}}$ in~\eqref{eq:channel1}. The IR decides that message index $j$, with $j \in \lbrace 1,2, \ldots, M \rbrace$, was transmitted, if the following event takes place:
\begin{equation}
    {\boldsymbol Y} \in \mathcal{D}_j,
\end{equation}
with $\mathcal{D}_j$ in~\eqref{Eqnm_code}.
That is, the set $\mathcal{D}_j \in \mathds{C}^n$ is the region of correct detection for message index $j$.
Therefore, the DEP associated with the transmission of message index $i$ is
\begin{IEEEeqnarray}{rCl}
\label{EqDEPi}
    \gamma_i(\mathscr{C}) 
    &\triangleq& 1 - \int_{\mathcal{D}_i} f_{\boldsymbol{Y}|\boldsymbol{X}}(\boldsymbol{y}|\boldsymbol{u}(i)) \mathrm{d}\boldsymbol{y},
\end{IEEEeqnarray}
and the average DEP is given by
\begin{IEEEeqnarray}{rCl} 
    \label{eq:gamma} 
    \gamma(\mathscr{C}) &\triangleq& \frac{1}{M}\sum_{i = 1}^M \gamma_i(\mathscr{C}). 
\end{IEEEeqnarray}

This leads to the following refinement of Definition~\ref{DefNmCode}.
\begin{definition}[$(n,M,\epsilon)$-codes]  
\label{def:nme_code}
An $(n,M)$-code $\mathscr{C}$ for the random transformation in~\eqref{EqChannelModel} is said to be an $(n,M,\epsilon)$-code  if 
\begin{equation}\label{EqGammaUpperbound}
    \gamma(\mathscr{C}) \leq \epsilon.
\end{equation}
\end{definition}
\subsection{Energy Transmission} \label{SubsecEnergyTransmission}
Given a channel output $z \in \mathds{C}$, the energy harvested from such channel output is given by a positive monotone increasing circularly symmetric function $g$ given by
\begin{equation} \label{EqEnergyFunction}
    g:\mathds{C}\rightarrow [0, +\infty),
\end{equation}
with $g(0) = 0$.
The energy transmission task must ensure that a minimum average energy $B$ is harvested at the EH at the end of $n$ channel uses. 
Let $\bar{g}:\mathds{C}^n \rightarrow [0, +\infty)$ be a positive function such that given $n$ channel outputs ${\boldsymbol z} = (z_1,z_2, \ldots, z_n)$, the average energy is
\begin{equation}
    \bar{g}({\boldsymbol z}) = \frac{1}{n}\sum_{t=1}^n g(z_t),
\end{equation}
in energy units per channel use.
Assume that the transmitter uses the code $\mathscr{C}$ in~\eqref{Eqnm_code}. Then, the EOP associated with the transmission of message index $i$, with $i \in \lbrace 1,2, \ldots, M \rbrace$, is 
\begin{IEEEeqnarray}{rCl}
\label{Eqthetai}
\theta_i(\mathscr{C},B) & \triangleq & \mathrm{Pr}[\bar{g}({\boldsymbol Z}) < B |{\boldsymbol X} = {\boldsymbol u}(i)], 
\end{IEEEeqnarray}
where the probability is with respect to the probability density function $f_{\boldsymbol{Z}|\boldsymbol{X}}$ in~\eqref{eq:z_distribution}; and the average EOP is given by
\begin{IEEEeqnarray}{rcl} 
\label{eq:theta_def}
    \theta(\mathscr{C},B) & \triangleq & \frac{1}{M}\sum_{i=1}^M \theta_i(\mathscr{C},B). 
\end{IEEEeqnarray}
This leads to the following refinement of Definition~\ref{def:nme_code}.
\begin{definition}[$(n,M,\epsilon,B,\delta)$-code] \label{def:nmed_code}
An $(n,M,\epsilon)$-code $\mathscr{C}$ for the random transformation in~\eqref{EqChannelModel} is said to be an $(n,M,\epsilon,B,\delta)$-code if 
\begin{equation} \label{eq:delta}
    \theta(\mathscr{C},B) \leq \delta.
\end{equation}
\end{definition}
The results in this work are presented in terms of the types~\cite{CsiszarMoT} induced by the codewords of a given code.
Given an $(n,M,\epsilon,B,\delta)$-code $\mathscr{C}$ of the form in~\eqref{Eqnm_code}, the type induced by the codeword $\boldsymbol{u}(i)$, with $i \inCountK{M}$, is a probability mass function (pmf) whose support is $\mathcal{X}$ in~\eqref{EqCIsymbols}. Such pmf is denoted by $P_{\boldsymbol{u}(i)}$ and for all $\ell \in \lbrace 1,2, \ldots, L \rbrace$,
\begin{equation} \label{eq:u_measure}
    P_{\boldsymbol{u}(i)}(x^{(\ell)}) \triangleq \frac{1}{n} \sum_{t=1}^n \mathds{1}_{\{u_t(i) = x^{(\ell)}\}},
\end{equation}
where $x^{(\ell)}$ is an element of $\mathcal{X}$ in~\eqref{EqCIsymbols}.
The type induced by all the codewords in $\mathscr{C}$ is also a pmf on the set $\mathcal{X}$ in~\eqref{EqCIsymbols}. Such pmf is denoted by  $P_{\mathscr{C}}$ and  for all $\ell \in \lbrace 1,2, \ldots, L \rbrace$,
\begin{equation} \label{eq:p_bar}
    P_{\mathscr{C}}(x^{(\ell)}) \triangleq \frac{1}{M} \sum_{i=1}^M P_{\boldsymbol{u}(i)}(x^{(\ell)}).
\end{equation}
A class of codes that is of particular interest in this study is that of  homogeneous codes, which are defined hereunder.
\begin{definition}[Homogeneous Codes]\label{DefHC}
An $(n,M,\epsilon,B,\delta)$-code $\mathscr{C}$ for the random transformation in~\eqref{EqChannelModel} of the form in~\eqref{Eqnm_code} is said to be homogeneous if for all $i \inCountK{M}$ and for all $\ell \inCountK{L}$, it holds that
\begin{equation}\label{EqHomogeneousCodes}
 P_{\boldsymbol{u}(i)}(x^{(\ell)}) =  P_{\mathscr{C}}(x^{(\ell)}),
\end{equation}
where, $P_{\boldsymbol{u}(i)}$ and $P_{\mathscr{C}}$ are the types defined in~\eqref{eq:u_measure} and~\eqref{eq:p_bar}, respectively.  
\end{definition}
Homogeneous codes are essentially $(n,M,\epsilon,B,\delta)$-codes that satisfy the condition that a given channel input symbol is used with the same frequency in all the codewords. Such codes are the primary focus of the results in this paper.
\section{Code Construction} \label{SecConstruction}
The information-energy capacity region of SIET systems with finite constellations is defined as follows.
\begin{definition}
The information-energy capacity region $\mathcal{C}(n,\epsilon,\delta)$ for the random transformation in~\eqref{EqChannelModel} is the set of all information and energy transmission rate pairs $(R,B) \in \reals^2$ for which there exists an $(n,M,\epsilon,B,\delta)$-code $\mathscr{C}$ such that $\frac{\log M}{n} = R$, the average DEP $\lambda(\mathscr{C}) \leq \epsilon$, and, the average EOP $\theta(\mathscr{C}) \leq \delta$.
\end{definition}
The process of characterizing an achievable information-energy region for SIET with finite constellations begins with the construction of an $\left( n,M \right)$-code.
Let the $\left( n,M \right)$-code $\mathscr{C}$ be 
\begin{subequations}
\label{EqnmCodeCircle}
\begin{equation} \label{Eq25a}
    \mathscr{C} \triangleq \left \lbrace ({\boldsymbol u}(1),\mathcal{D}_1), ({\boldsymbol u}(2),\mathcal{D}_2), \ldots, ({\boldsymbol u}(M),\mathcal{D}_M)\right \rbrace.
\end{equation}

The construction of the code begins with the construction of the channel input symbols. The set of channel input symbols is a modulation constellation represented by a finite subset of $\complex$.
Consider a constellation formed by $C$ \emph{layers}, with $C \in \ints$. A layer is a subset of symbols in $\complex$ that have the same magnitude. For all $c \in \lbrace 1,2, \ldots, C \rbrace$, denote by  $L_c \in \ints$ the number of symbols in the $c$\ts{th} layer and let $A_c \in \reals^+$ be the amplitude of the symbols in layer $c$. Denote such a layer by $\mathcal{U}(A_c,L_c)$. That is, 
\begin{IEEEeqnarray}{l}
    \label{EqLayerCircle}
    \mathcal{U}(A_c,L_c) \triangleq
    \Big\lbrace x_c^{\left( \ell \right)} = \\
    A_c \exp\left(\mathrm{i} \frac{2\pi}{L_c} \ell \right) \subseteq \complex: \ell \in \left \lbrace 0, 1,2, \ldots, (L_c-1)\right \rbrace \Big\rbrace, \nonumber
\end{IEEEeqnarray} 
where $\mathrm{i}$ is the complex unit.
Using this notation, the constellation can be described by the following set:
\begin{IEEEeqnarray}{rCl} \label{EqConstellationCircle}
    \mathcal{X} & = & \bigcup_{c=1}^C \mathcal{U}(A_c,L_c).
\end{IEEEeqnarray}
%
Without any loss of generality, assume that
\begin{equation} \label{EqAmplitudesOrder}
    A_1 > A_2 > \ldots > A_C.
\end{equation}
The symbols in layer $c$ of the form in~\eqref{EqLayerCircle}, are equally spaced along a circle of radius $A_c$. The constellation induced by the set $\mathcal{X}$ is thus made up of points uniformly distributed along $C$ concentric circles. 
%
The total number of symbols $L$ in~\eqref{EqL} for $\mathcal{X}$ in~\eqref{EqConstellationCircle} is 
\begin{equation} \label{EqLSum}
    L = \sum_{c=1}^C L_c.
\end{equation}

The construction of the $(n,M)$-code $\mathscr{C}$ in~\eqref{Eq25a} is as follows.
For all $c \inCountK{C}$, let $p_c$ be the frequency with which symbols of the $c$\ts{th} layer appear in the code. The resulting probability vector is denoted by
\begin{equation} \label{Eqp}
    \boldsymbol{p} = \left( p_1,p_2, \ldots, p_C \right)^{\sf{T}},
\end{equation}
where, for all $c \inCountK{C}$,
\begin{equation} \label{Eqpc}
    p_c = \frac{1}{Mn} \sum_{\ell = 1}^{L_c} \sum_{i=1}^M \sum_{t=1}^n \mathds{1}_{\{u_t(i) = x_c^{(\ell)}\}}.
\end{equation}
\par
The decoding set $\mathcal{G}_c^{(\ell)}$ associated with symbol $x_c^{(\ell)}$ is a circle of radius $r_c \in \reals^+$ centered at $x_c^{(\ell)}$. That is,
\begin{equation} \label{EqDecodingCircle}
\mathcal{G}_c^{(\ell)} = \left\lbrace y \in \complex : \left| y - x_c^{\left(\ell \right)}\right|^2 \leq r_c^2 \right\rbrace. 
\end{equation}
The radii $r_1$, $r_2$, $\ldots$, $r_C$ are chosen such that the decoding regions are mutually disjoint. To ensure this, for all $c \inCountK{C}$ the amplitudes $A_c$ in~\eqref{EqLayerCircle} satisfy the following
\begin{IEEEeqnarray}{rcl} 
\label{EqAmplitudesDifference}
A_c - A_{c-1} \geq r_c + r_{c-1}.
\end{IEEEeqnarray}
The vector formed by these radii is denoted by
\begin{equation} \label{EqRadiusVector}
\boldsymbol{r} = \left( r_1, r_2, \ldots, r_C\right)^{\sf{T}}.
\end{equation}
For all $i \inCountK{M}$, the decoding region for codeword $\boldsymbol{u}(i)$ is
\begin{equation} \label{EqDecodingCodeword}
    \mathcal{D}_i = \mathcal{D}_{i,1} \times \mathcal{D}_{i,2} \times \ldots \times \mathcal{D}_{i,n}, 
\end{equation}
\end{subequations}
where $c \inCountK{C}$, $t \inCountK{n}$, and $\ell \inCountK{L_c}$ are such that if, $u_t(i) = x_c^{(\ell)}$, then, $\mathcal{D}_{i,t} = \mathcal{G}_c^{(\ell)}$.

This defines a family of $(n,M)$-codes denoted by
\begin{IEEEeqnarray}{rCl}
\label{EqFamily}
\sf{C\left(\mathcal{X},\boldsymbol{p},\boldsymbol{r} \right)},
\end{IEEEeqnarray}
with constellation $\mathcal{X}$ in~\eqref{EqConstellationCircle}, probability vector $\boldsymbol{p}$ in~\eqref{Eqp} and, radii of decoding regions $\boldsymbol{r}$ in~\eqref{EqRadiusVector}.
\section{Converse Results} \label{SecConverse}
This section characterizes a converse region for codes in $\sf{C\left(\mathcal{X},\boldsymbol{p},\boldsymbol{r} \right)}$ in~\eqref{EqFamily}. That is, any tuple $(n,M,\epsilon,B,\delta)$ outside the converse region is not achievable by the codes in $\sf{C\left(\mathcal{X},\boldsymbol{p},\boldsymbol{r} \right)}$. 

The following theorem follows from Lemma $1$ and Lemma $3$ in~\cite{zuhraITW} and Lemma $4.7$ in~\cite{zuhra}.
\begin{theorem}\label{TheoremConverse}
Consider a homogeneous $(n,M,\epsilon,B,\delta)$-code $\mathscr{C}$ for the random transformation in~\eqref{EqChannelModel} with constellation $\mathcal{X}$ of the form in~\eqref{EqConstellationCircle} and $\boldsymbol{p} = \left( p_1,p_2, \ldots, p_C \right)^{\sf{T}}$ in~\eqref{Eqp}. The constellation $\mathcal{X}$ has $C$ layers and for all $c \inCountK{C}$, layer $c$ contains $L_c$ symbols and has amplitude $A_c$. Then, the following holds
\begin{subequations}
\begin{IEEEeqnarray}{l}
M \leq  \frac{n!}{\prod_{\ell=1}^L (n\frac{p_c}{L_c})!}; \label{EqConverseR} \\
\label{EqConverseB}
B \leq \frac{1}{1-\delta} \sum_{c=1}^C p_c \mathbb{E}_{\sf{W}}\left[g\left(A_c + \sf{W} \right) \right]; \mbox{ and }\\
\nonumber
    \epsilon \geq 1 - \prod_{c=1}^C \prod_{\ell=1}^{L_c} \Bigg(1 - Q \Bigg( \frac{ |x_c^{(\ell)} - \bar{x}_{c}^{(\ell)}|}{\sqrt{2\sigma^2}}  \\
    - \frac{\sigma}{\sqrt{2} |x_c^{(\ell)} - \bar{x}_{c}^{(\ell)}|} \log \left( \frac{P_{\mathscr{C}}(\bar{x}_{c}^{(\ell)})}{P_{\mathscr{C}}(x_c^{(\ell)})} \right) \Bigg) \Bigg)^{n \frac{p_c}{L_c}}, \label{EqConversee}
\end{IEEEeqnarray}
\end{subequations}
where, the type $P_{\mathscr{C}}$ is defined in~\eqref{eq:p_bar}; the function $g$ is the energy function in~\eqref{EqEnergyFunction}; 
the expectation in~\eqref{EqConverseB} is with respect to $\sf{W}$, which  is a complex circularly symmetric Gaussian random variable whose real and imaginary parts have zero means and variances $\frac{1}{2}\sigma^2$; and,
for all $c \inCountK{C}$ and $\ell \inCountK{L}$, the complex $\bar{x}_{c}^{(\ell)}$ satisfies
\begin{IEEEeqnarray}{l}
\nonumber
    \bar{x}_{c}^{(\ell)} \in \argm_{x \in \mathcal{X} \setminus \{x_c^{(\ell)} \} }  \Bigg(1 - \\
\label{EqQfunction}    Q \Bigg( \frac{|x_c^{(\ell)} - x|}{\sqrt{2\sigma^2}} - \frac{\sigma}{\sqrt{2} |x_c^{(\ell)} - x|} \log \left( \frac{P_{\mathscr{C}}(x)}{P_{\mathscr{C}}(x_c^{(\ell)})} \right) \Bigg) \Bigg). 
\end{IEEEeqnarray}
The function $Q$ in~\eqref{EqConversee} and~\eqref{EqQfunction} is the $Q$ function defined in~\cite[Chapter $2$]{proakis}.
\end{theorem}

\section{Main Results} \label{SecMain}
This section provides various achievability results for homogeneous codes in the family $\sf{C\left(\mathcal{X},\boldsymbol{p},\boldsymbol{r} \right)}$. The proofs of these Lemmas are presented in~\cite{zuhra}.

\subsection{Information Transmission}
The results in this subsection provide conditions on the parameters of homogeneous codes in $\sf{C\left(\mathcal{X},\boldsymbol{p},\boldsymbol{r} \right)}$ that impact the information transmission rate $R$ and the DEP $\epsilon$.
\begin{lemma} \label{CorRadiusHomogeneous}
Consider a homogeneous $\left( n,M \right)$-code $\mathscr{C}$ for the random transformation in~\eqref{EqChannelModel} of the form in~\eqref{EqnmCodeCircle} with $\boldsymbol{p} = \left( p_1,p_2, \ldots, p_C \right)^{\sf{T}}$ in~\eqref{Eqp}. The code $\mathscr{C}$ is an $\left( n,M,\epsilon \right)$-code if the parameters $r_1, r_2, \ldots, r_C$ in~\eqref{EqDecodingCircle} satisfy
\begin{equation} \label{EqTheoremrc}
    \prod_{c=1}^{C}  \left( 1-e^{-\frac{r_c^2}{\sigma^2}} \right)^{n p_c} \geq 1-\epsilon,
\end{equation}
where, the real $\sigma^2$ is defined in~\eqref{EqDensities}.
\end{lemma}
%
%
\begin{lemma} \label{LemmaLc}
Consider an $\left( n,M \right)$-code $\mathscr{C}$ for the random transformation in~\eqref{EqChannelModel} of the form in~\eqref{EqnmCodeCircle}, with the set of symbols $\mathcal{X}$ in~\eqref{EqConstellationCircle} and $\boldsymbol{p} = \left( p_1,p_2, \ldots, p_C \right)^{\sf{T}}$ in~\eqref{Eqp}. Then, for all $c \inCountK{C}$, the number of symbols in layer $c$ of $\mathcal{X}$ is given by
\begin{equation} \label{EqTheoremLc}
    L_c \leq \left\lfloor \frac{\pi}{2\arcsin{\frac{r_c}{2A_c}}} \right\rfloor,
\end{equation}
and, the number of codewords $M$ satisfies the following:
\begin{IEEEeqnarray}{rcl}
\label{EqTheoremR}
M \leq  \frac{n!}{\prod_{c=1}^C \left((n\frac{ p_c}{L_c})! \right)^{L_c}},
\end{IEEEeqnarray}
where, $r_c$ is the radius of the decoding regions $\mathcal{G}_c^{(1)}, \ldots, \mathcal{G}_c^{(L_c)}$ in~\eqref{EqDecodingCircle} and $A_c$ is the amplitude in~\eqref{EqLayerCircle}.
\end{lemma}
%
%
\subsection{Energy Transmission}
The result in this subsection provides conditions on the parameters that impact the energy transmission rate $B$ and the EOP $\delta$ for homogeneous codes in $\sf{C\left(\mathcal{X},\boldsymbol{p},\boldsymbol{r} \right)}$.
\begin{lemma} \label{CorollaryAmplitudeC}
Consider a homogeneous $\left( n,M, \epsilon \right)$-code $\mathscr{C}$ for the random transformation in~\eqref{EqChannelModel} of the form in~\eqref{EqnmCodeCircle} with $\boldsymbol{p} = \left( p_1,p_2, \ldots, p_C \right)^{\sf{T}}$ in~\eqref{Eqp}. The code $\mathscr{C}$ is an $\left( n,M,\epsilon,B,\delta \right)$-code if, the energy transmission rate $B$ satisfies the following:
\begin{equation} \label{EqTheoremB}
B \leq \frac{1}{1-\delta} \sum_{c=1}^C p_c \mathbb{E}_{\sf{W}}\left[g\left(A_c + \sf{W} \right) \right]
\end{equation}
where, the parameters $A_c$ are in~\eqref{EqLayerCircle} and, the expectation is with respect to $\sf{W}$, which is a complex circularly symmetric Gaussian random variable whose real and imaginary parts have zero means and variances $\frac{1}{2}\sigma^2$.
\end{lemma}
%
%
\begin{figure}[t]
  \centering
  \includegraphics[width=0.5\textwidth]{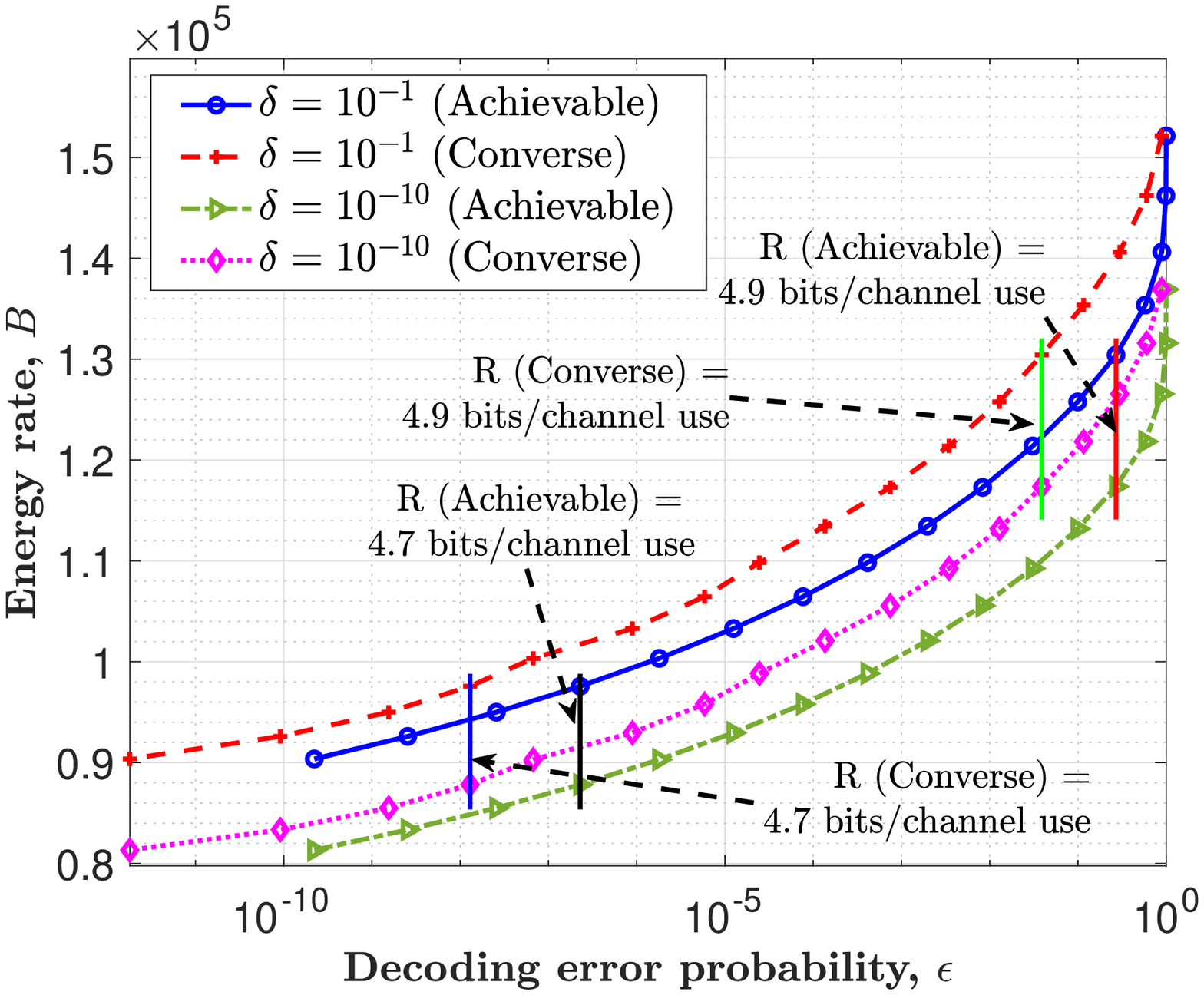}
  \caption{Converse~\eqref{EqConverseB} and achievable~\eqref{EqTheoremB} bounds on energy transmission rate $B$ for a homogeneous code in the family $\sf{C\left(\mathcal{X},\boldsymbol{p},\boldsymbol{r} \right)}$ as a function of the DEP $\epsilon$ in~\eqref{EqConversee} and~\eqref{EqTheoremrc}, respectively.}
\label{FigBDEP}
\end{figure}
\section{Final Remarks} \label{SecExample}
Consider a homogeneous $(n,M,\epsilon,B,\delta)$-code $\mathscr{C}$ in $\sf{C\left(\mathcal{X},\boldsymbol{p},\boldsymbol{r} \right)}$ in~\eqref{EqFamily}. The constellation $\mathcal{X}$ is of the form in~\eqref{EqConstellationCircle} with number of layers $C = 3$. Duration of the transmission in channel uses is $n = 80$. The energy harvested at the EH takes into consideration the  non-linearities of the receiver as suggested in~\cite{8115220,varasteh2017wireless}. More specifically, the energy function $g$ in~\eqref{EqEnergyFunction} is of the form in~\cite[Proposition $1$]{varasteh2017wireless}.
\par
In Fig.~\ref{FigBDEP}, the bound on the achievable energy transmission rate $B$ in~\eqref{EqTheoremB} for code $\mathscr{C}$ is plotted as a function of the achievable DEP $\epsilon$ in~\eqref{EqTheoremrc}. The figure also shows the converse bound on $B$ in~\eqref{EqConverseB} as a function of $\epsilon$ in~\eqref{EqConversee}.
The amplitude of the first layer is $A_1 = 50$. Amplitudes of the second and third layers $A_2$ and $A_3$ are determined by the radii of the decoding regions according to~\eqref{EqAmplitudesDifference}. For all $c \inCountK{C}$, the number of symbols in layer $c$ i.e., $L_c$, is determined by the radii $r_c$ and the amplitudes $A_c$ according to~\eqref{EqTheoremLc}. The probability vector in~\eqref{Eqp} is $\boldsymbol{p} = (0.5,0.3,0.2)^{\sf{T}}$. The points on the curves are generated by varying $r_c$ between $2$ and $10$.
\par
Fig.~\ref{FigBDEP} shows several trade-offs between the energy transmission rate $B$, the DEP $\epsilon$, the EOP $\delta$, and the information transmission rate $R$. Firstly, the energy rate $B$ increases as $\epsilon$ increases. This effect is due to the fact that increasing $\epsilon$ allows decreasing the radii of the decoding regions $\boldsymbol{r}$ in~\eqref{EqRadiusVector} according to~\eqref{EqTheoremrc}. At the same time, decrease in $r_c$ allows increasing the amplitudes $A_2$ and $A_3$ according to~\eqref{EqAmplitudesDifference} which increases $B$ according to~\eqref{EqTheoremB}.
Secondly, the energy rate $B$ increases as $\delta$ increases. This effect stems from the dependence of $B$ on $\delta$ as in~\eqref{EqTheoremB}.
Thirdly, the information rate $R$ increases as $\epsilon$ increases. This is because increasing $\epsilon$ allows decreasing $r_c$ according to~\eqref{EqTheoremrc}. At the same time, decrease in $r_c$ allows increasing the number of symbols in a layer $L_c$ according to~\eqref{EqTheoremLc} which increases $R$ according to~\eqref{EqTheoremR} and~\eqref{EqR}.
\begin{figure}[t]
  \centering
  \includegraphics[width=0.5\textwidth]{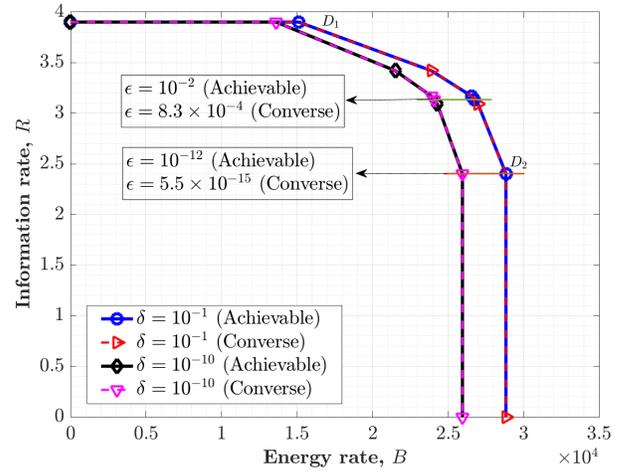}
  \caption{Converse and achievable information-energy regions for homogeneous codes in the family $\sf{C\left(\mathcal{X},\boldsymbol{p},\boldsymbol{r} \right)}$.}
\label{FigRegions}
\end{figure}

Fig.~\ref{FigRegions} shows the converse and achievable information-energy regions of code $\mathscr{C}$ as a function of the EOP $\delta$ and the DEP $\epsilon$. The radii of the decoding regions $r_c$ are assumed to be the same for all the layers i.e., for all $c \inCountK{C}$, the radii $r_c = r$ in~\eqref{EqDecodingCircle}. The value of $r$ is obtained according to~\eqref{EqTheoremrc} to satisfy $\epsilon$. The amplitude of the first layer is $A_1 = 30$. Amplitudes of the second and third layers $A_2$ and $A_3$ are determined by $r$ according to~\eqref{EqAmplitudesDifference}. The points in Fig.~\ref{FigRegions} are obtained by varying $\epsilon$ and the probability vector $\boldsymbol{p}$ in~\eqref{Eqp}.
\par
Fig.~\ref{FigRegions} shows the following trade-offs between the information and energy transmission rates in the converse and achievable curves.
Firstly, the maximum achievable information transmission rate is $R = 3.9$ bits/channel use. This $R$ is achieved by a code in which all the symbols in the constellation $\mathcal{X}$ are used with the same frequency. The maximum energy transmission rate that can be achieved at $R = 3.9$ bits/channel use is $B = 1.5 \times 10^4$ energy units. This corresponds to the point $D_1$ in Fig.~\ref{FigRegions}.
Secondly, the maximum achievable $B$ is $2.9 \times 10^{4}$ energy units. This is achieved by a code that exclusively uses the symbols in the first layer i.e., the probability vector $\boldsymbol{p}$ in~\eqref{Eqp} is $\boldsymbol{p} = (1,0,0)^{\sf{T}}$. The maximum $R$ that can be achieved at $B = 2.9 \times 10^{4}$ energy units is $R = 2.4$ bits/channel use. This corresponds to the point $D_2$ in Fig.~\ref{FigRegions}.
Thirdly, the curves between the points $D_1$ and $D_2$ in Fig.~\ref{FigRegions} show the trade-off between the information and energy transmission rates. As $B$ is increased from $1.5 \times 10^4$ energy units at point $D_1$, $R$ decreases. Similarly, as $R$ is increased from $2.4$ bits/channel use at point $D_2$, $B$ decreases.
\subsection{Comments on Optimality}
The codes constructed in this work are optimal in the sense of the converse results of Theorem~\ref{TheoremConverse} except for the DEP $\epsilon$. Fig.\ref{FigBDEP} shows that the code $\mathscr{C}$ achieves the optimal energy rate $B$ and information rate $R$ as given by the converse results albeit at a higher DEP $\epsilon$. Fig.~\ref{FigRegions} shows that the converse and achievable information-energy rate curves for $\mathscr{C}$ overlap. However, for the same information and energy rate pair, the DEP for the achievable curves is higher than that of the converse curves. The sub-optimality in DEP arises due to the sub-optimal choice of circular decoding regions in~\eqref{EqDecodingCircle}. 

The proposed construction provides a method of building codes that meet the given energy and information rate, EOP, and DEP requirements. Building codes that achieve the optimal energy and information rate, EOP, and DEP requires optimizing set of channel input symbols $\mathcal{X}$. However, the problem of optimal input design even for the most well behaved channels~\cite{8613368,8878162,8849318} remains an open problem. 






\bibliographystyle{IEEEtran}
\bibliography{myrefs}
\end{document}